\documentclass[12pt,a4paper,twoside]{article}

\def\TheAuthor{Maria Hermanns}                                   
\def\TheTitle{Entanglement in topological systems}                        
\def\TheHeading{Entanglement in Topological Systems}                                
\def\TheInstitute{Institute for Theoretical Physics}              
\def\TheAddress{University of Cologne}            
\def\ThePart{D}                                             
\def\TheChapter{6}                                          

\usepackage{ferienschule_15}                                  
\usepackage{color}

\makeindex

\begin{document}

\MakeTitel           
\tableofcontents     

\vfill
\rule{42mm}{0.5pt}\\
{\footnotesize Lecture Notes of the $48^{{\rm th}}$ IFF Spring
School ``Topological Matter -- Topological Insulators, Skyrmions \newline and Majoranas''
(Forschungszentrum J{\"{u}}lich, 2017). All rights reserved. }

\newpage


\section{Introduction}

The way we understand and describe condensed matter systems has shifted quite substantially in the last few decades, a development that  started with the discovery of the quantum Hall effect \cite{IQH, FQH}. Previously, it was believed that Landau's symmetry breaking paradigm could classify all phases of matter. It not only gave a comprehensive classification of crystals and other conventionally ordered phases, but also of a plethora of superfluid and superconducting phases. It  also explained the rather surprising universality arising in continuous phase transitions. It was due to this success that the discovery of the quantum Hall effect had such a tremendous impact on the field. Suddenly, there were distinct phases of matter that could not be distinguished by local order parameters. It became apparent that the Landau paradigm of broken symmetries was incomplete, and that new theoretical tools had to be invented. 

How do we classify phases of matter, if symmetry is not sufficient? Condensed matter physics has imported concepts from very different research areas to come up with an answer to this question. One of the most central new concepts is \emph{topology}. Its first appearance in this context was in the seminal paper by Thouless, Kohmoto, Nightingale, and den Nijs, where they related the quantized Hall conductance of the integer quantum Hall effect to a \emph{topological invariant} -- the Chern number \cite{Tho82}. Topological invariants have now become the standard tool for classifying non-interacting topological phases \cite{Classification1,Classification2,Classification3}. This field has grown rapidly in the last years, mainly due to a fruitful interplay between theoretical predictions and experimental discoveries, e.g. in the context of topological insulators \cite{Kan05,Kan05+,Hsi08}  or Weyl semimetals \cite{Wan11,Xu15}. For an overview of the field, see e.g. the lecture notes by Fritz and Altland \cite{PrimerTI}, the review by Hasan and Kane \cite{HasanKane}, or the book by Bernevig \cite{BernevigBook}. 

The other important concept, imported to condensed matter physics from quantum information, is \emph{entanglement}. 
Strongly interacting phases, such as the fractional quantum Hall liquids, are \emph{topologically ordered} \cite{topologicalorder} or equivalently \emph{long-range entangled} \cite{QImeetsCondMat}. Even though such phases are gapped, and all correlation functions decay exponentially,  the entanglement between different parts of the system can persist to arbitrary long distances.   

In order to detect long-range entanglement one has to develop ``non-local measures". In these lectures, we are going to discuss two such measures --- the entanglement entropy and the entanglement spectrum. The discussion focuses  on two-dimensional, topologically ordered systems, but there are many other systems where the entanglement entropy/spectrum is used as a tool. These lectures will not give a comprehensive overview of the various areas of applications, and will only provide a selected list of references at the end for the interested reader.

\subsection{Topological phases of matter}

Most of our understanding of topological phases is based on the study of idealized models. This has been  successful because topological properties are extremely robust and do not depend on microscopic details. 
However, there are complications that do matter. In particular, there is an important distinction between those phases that can qualitatively be understood using models of noninteracting particles and those where interactions are essential. 
Interactions allow for the fractionalization of quantum numbers and can lead to the emergence of \emph{anyonic excitations} that are fundamentally different from elementary particles. 
Such phases of matter are called \emph{topologically ordered}\iffindex{topological order}, and have various special properties that distinguish them from conventionally ordered states. Their effective low-energy theory is described by a topological quantum field theory (TQFT), which encodes the universal (topological) behavior at long distances. The physics at short distances, on the other hand, is not universal and depends crucially on the details of the system. 

Arguably one of the  most fascinating features of topologically ordered phases is that they harbor anyonic quasiparticle excitations \cite{anyons, anyons2}. These are particle-like collective excitations that can only occur in two-dimensions. Upon exchanging two \emph{abelian} anyons, the wave function acquires a fractional phase factor $e^{i\alpha}$, where $\alpha$ may take any value between 0 (bosons) and $\pi$ (fermions).\footnote{On a compact manifold, such as the sphere and the torus, $\alpha$ needs to be a fraction in order to yield a well-defined theory, but on a non-compact manifold such as the plane, truly any value of $\alpha$ is allowed.}
\emph{Nonabelian} anyons exhibit an even more exotic behavior \cite{MooreRead}. For these, the many-anyon state is not uniquely determined by the positions of the anyons. Instead there are several (degenerate) states that form a vector space, and the wave function is an (arbitrary) vector in this space. Exchanging two anyons is implemented by a (unitary) rotation, which does not depend on the details of the exchange path but only on its topology. Nonabelian anyons have, therefore, been proposed as a very robust way to implement quantum gates and build a topologically protected quantum computer \cite{TopQuComp}. A more detailed discussion on this can be found in the lectures on \emph{Topological quantum computing} by David DiVincenzo.    
An alternative way to define nonabelian anyons is via their \emph{quantum dimension}. The quantum dimension $d_i$ of an anyon with charge $i$ is a measure how the Hilbert space of localized anyons grows asymptotically, i.e. dim$(\mathcal H) \propto d_i^n$ for $n$ anyons. Abelian anyons have quantum dimension $d_i=1$, nonabelian anyons have a quantum dimension $d_i>1$.
 This is intuitively clear, as the wave function of abelian anyons is uniquely specified (up to a phase) by the anyon position. Thus, the Hilbert space dimension for any number of abelian anyons is one.  The  wave function of nonabelian anyons is, however, not uniquely specified by their position, and the Hilbert space dimension grows with the number of anyons. 
 As an example, let us consider Majorana fermions which can be interpreted as `half' a fermion. $2n$ Majorana fermions form $n$ complex fermionic modes,  each of which can be occupied or empty. 
Therefore, the Hilbert space of $2n$ (localized) Majorana fermions has size $2^n=\sqrt{2}^{2n}$ (or $2^{n-1}$ for each parity sector) and the corresponding quantum dimension of a Majorana fermion is $\sqrt{2}$. Note that quantum dimensions need not be square roots of integers -- e.g. the quantum dimension of the so-called Fibonacci anyons is the golden ratio $\phi=(1+\sqrt{5})/2$ \cite{Fibonacci}.

Topologically ordered phases have (in the thermodynamic limit) a ground state degeneracy that depends on the genus (i.e. the number of handles) of the manifold it is placed on. On a torus, which has genus 1, the ground state degeneracy is given by the number of distinct quasiparticles types $n_{qp}$ in the corresponding TQFT. For higher genus $g$, the degeneracy is $n_{qp}^g$. 

 Last but not least, topologically ordered states are  \emph{long-range entangled} \cite{LRE}\iffindex{long-range entanglement}. 
 A state is long-range entangled, if it cannot be transformed to a product state under generalized stochastic linear transformations \cite{LRE}, otherwise it is called short-range entangled. We will not go into further details on the precise definition of short- and long-range entanglement, but rather refer the reader to the book by Wen et al. \cite{QImeetsCondMat} for a pedagogical review on this field. In these lecture notes, it is sufficient to know that long-range entanglement is equivalent to a non-zero value of the \emph{topological entanglement entropy} $\gamma$ \cite{EEtopological1,EEtopological2}. The latter is  the logarithm of the \emph{total quantum dimension}\iffindex{total quantum dimension} $\mathcal D=\sqrt{\sum d_i^2}$ of the TQFT.

While there are many interesting theoretical proposals for topologically ordered phases, there are very few experimental realizations. In fact,  the only examples that have been experimentally verified beyond any doubt are the fractional quantum Hall liquids \cite{FQH}. These are believed to realize mostly abelian phases, but some fractions in the first excited Landau level are proposed to harbor nonabelian excitations \cite{MooreRead}. There are many potential candidates for topologically ordered states in the context of quantum spin liquids.\footnote{See for instance the review by Balents \cite{BalentsReview} for further details, as well as the discussion of Kitaev spin liquids \cite{KSL}  in the lectures by Simon Trebst.} Quantum spin liquids are magnetic systems where the spins fluctuate strongly even at zero temperature and consequently  no magnetic order can develop. While several experiments are certainly promising, there is still no universally accepted experimental demonstration of a true quantum spin liquid ground state in experiments. 
One of the main reasons for this is that long-range entanglement -- the hallmark of topological order -- has so far only been `measured' in numerics.\footnote{This is not entirely true any longer --- the group of Greiner measured the second R\'enyi entropy in a cold atom system \cite{EEexperiment}, but it is certainly true for condensed matter systems.} There are, however, several theoretical proposals of how to measure entanglement in various setups, see for instance the proposals in Refs.~\cite{EntanglementProposals1,EntanglementProposals2,EntanglementProposals3}.

\subsection{The toric code} 
\iffindex{toric code}

\begin{figure}
    \centering
     \includegraphics[width=\hsize]{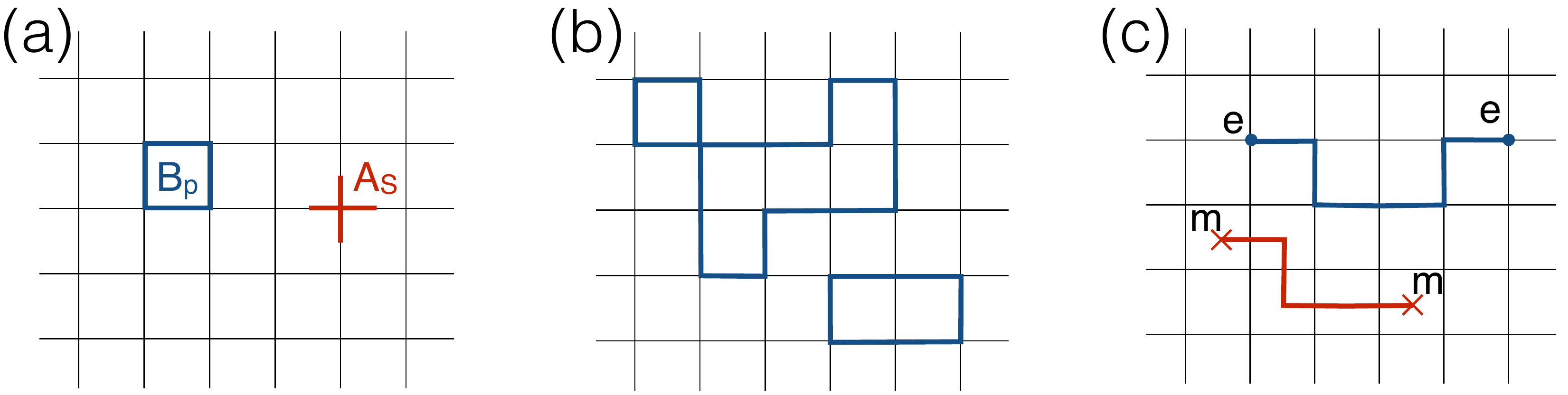}
     \caption{(a) Graphic representation of the star and plaquette operators. A red (blue) bond implies that $\sigma^z$ ($\sigma^x$) acts on the spin on the bond. (b) Example of an allowed loop configuration. (c) Graphic representation of the magnetic and electric path operators. }\label{fig:TC}
\end{figure} 
Here, we give a short introduction to the toric code \cite{toriccode}, as it is one of the simplest models of topologically ordered phases, and will be used as a example later in these lectures. 
It describes a gapped quantum spin liquid with fractionalized excitations that obey nontrivial mutual braiding statistics. It was originally proposed by Kitaev as a way to achieve fault-tolerant quantum computing.
We will restrict our discussion to the bare essentials that are needed later for the discussion on the entanglement entropy and entanglement spectrum. 
The interested reader is referred to the original article \cite{toriccode} or to the lecture notes by Kitaev and Laumann \cite{LaumannKitaev} for a more thorough discussion of this model. 

The toric code is an exactly solvable spin model, where spin 1/2 degrees of freedom are sitting on the edges of a square lattice. 
The Hamiltonian of the system is given by 
\begin{align}\label{eq:ToricCodeH}
 H &=-J_A\sum_{s} A_s -J_B\sum_{p} B_p \equiv H_A+H_B
\end{align}
with 
\begin{align}
A_s&=\prod_{j\in s} \sigma_j^z ,&
B_p&=\prod_{j\in p} \sigma_j^x
\end{align}
where the `star operator' $A_s$ contains the four spins around a vertex $s$ and the `plaquette operator' $B_p$ the four spins around a plaquette, see Fig.~\ref{fig:TC} (a). 
We assume that both $J_A$ and $J_B$ are positive. 

Let us first note that all the star and plaquette operators are mutually commuting $[A_s, A_{s'}]=[B_p,B_{p'}]=[A_s,B_p]=0$, and can be simultaneously diagonalized. 
They have eigenvalues $\pm 1$, and the ground state satisfies $A_s=+1$ and $B_p=+1$ for all stars and plaquettes. 
In the following, we will use the $\sigma^z$ eigenbasis.
Requiring $A_s=+1$ implies that  there is an even number of spin-ups around the vertex $s$. 
We can represent this graphically by putting a string on each link with a spin-$\downarrow$. 
The states that minimize $H_A$  are closed loop configurations, as exemplified in Fig.~\ref{fig:TC}(b). An open string, on the other hand, comes with two excited star operators, $A_s=-1$,  at its end points, see Fig.~\ref{fig:TC}(c). 
Such excited states of the toric code are discussed later in more detail. 

\begin{figure}
    \centering
     \includegraphics[width=\hsize]{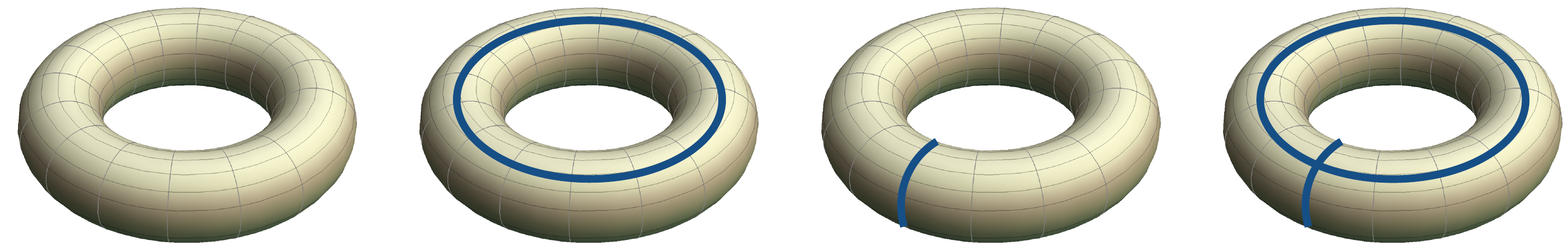}
     \caption{On the torus, there are four distinct sectors that are labeled by the number of loops mod~2 around each of the handles. Each picture depicts a  reference configuration, from which the respective ground state can be built by acting with $\prod_{j=1}^{N-1}(1+B_p)$.  }\label{fig:torus}
\end{figure} 

$H_A$ has a macroscopic degeneracy of $\sqrt{2}^{ \#(spins)}$, which is lifted by $H_B$. 
A plaquette operator $B_p$  flips all spins around a plaquette. Graphically, it creates (or destroys) a loop around plaquette $p$. Thus, acting with $H_B$ mixes states with different loop configurations. 
The ground state of the system is an equal weight superposition of all allowed loop configurations. 
On the plane, we can construct this superposition by the operator 
\begin{align}\label{eq:loopsum}
\prod_{p}(1+B_p)
\end{align} acting on a reference state, e.g. the state where all spins are $\uparrow$.
This is not the case on the torus. First, we notice that the periodic boundary conditions imply that $\prod_p B_p=1$ and  $\prod_s A_s=1$.
Consequently, the operator \eqref{eq:loopsum} generates each loop configuration twice. 
In addition, not all allowed loop configurations on the torus can be obtained by acting with products of $B_p$'s on a single reference configuration --- examples are the loop configurations shown in Fig.~\ref{fig:torus}. 
To make this more precise, let us introduce Wilson loop operators
\begin{align}\label{eq:Wilson}
W_{\ell}&=\prod_{j\in \ell} \sigma_j^x, 
\end{align}
 which are defined for any closed path $\ell$ on the lattice. It is straightforward to verify that they commute with each other as well as with the Hamiltonian. 
 If $\ell$ wraps around one of the handles of the torus, the loop is called non-contractible and cannot be represented as a product of plaquette operators. 
 On the torus, there are four distinct sectors, corresponding to the number mod~2 of non-contractible loops around each handle. 
 The ground state in each sector is obtained by acting with \eqref{eq:loopsum} on one of the reference configurations that are depicted graphically in Fig.~\ref{fig:torus}.
 A special feature of the toric code is that the ground states are exactly degenerate, even for a finite system size.

The ground state degeneracy is intimately related to the excitations of the system. 
We will refer to a site with $A_s=-1$ as carrying an `electric charge'. A pair of electric charges is created by  the electric path operator  
\begin{align}
W_\gamma^{(e)}&=\prod_{j\in\gamma} \sigma_j^x
\end{align}
where $\gamma $ is an open path on the lattice, and the electric charges are sitting at its end points, see  Fig.~\ref{fig:TC}(c). 
As the energy cost of $W_\gamma$ is $4J_A$ \emph{independent}  of the length of the path $\gamma$, we can separate the electric charges arbitrarily far -- they are \emph{deconfined}. 
Note that the Wilson loop operators \eqref{eq:Wilson} are nothing but the closed version of the $W_\gamma^{(e)}$. 
Thus, we can change the ground state sector by creating two electric charges, moving one around one (or both) of the handles of the torus, and afterwards annihilating them again. 
 
Similarly to the electric charges, we can define `magnetic' charges that are located on plaquettes with $B_p=-1$. A pair of magnetic charges is created by  the magnetic path operator
 \begin{align}
W_{\tilde \gamma}^{(m)}&=\prod_{j\in\tilde\gamma} \sigma_j^z
\end{align}
where $\tilde\gamma$ now lives on the dual lattice, see Fig.~\ref{fig:TC}(c). 
Analogously to the electric excitations, the energy cost for a pair of magnetic excitations is $4J_B$ independent on their distance --  magnetic excitations are deconfined. 

Both magnetic and electric charges are bosons. However, they have nontrivial mutual statistics. 
Braiding an electric charge around a magnetic one yields a negative sign in the wave function\footnote{The mutual sign can easiest be understood by realizing that  an electric  charge always drags a string of $\sigma_x$'s behind. 
Thus, the initial state and the final state (after having moved  the electric charge around the magnetic charge) differ by a Wilson loop operator \eqref{eq:Wilson} that encloses the magnetic charge. 
Since a contractible Wilson loop operator can be expressed as the product of enclosed plaquette operators, its eigenvalue is -1 (there is exactly one plaquette with eigenvalue -1 within the loop). }, similar to what we know from moving an elementary charge around a flux quantum in a superconductor. 
The important difference here is that the magnetic and electric charges in the toric code are fully equivalent. 
This becomes particularly easy to see when using Wen's version of the toric code, where both the star and the plaquette operators are mapped to (identical) plaquette operators \cite{WensTC}. 

Let us conclude this discussion by summarizing the topological properties of the toric code. There are four particle types in the theory: $\mathbf 1$ (the identity), the electric charge $\mathbf e$, the magnetic charge $\mathbf m$, and a fermion $\boldsymbol\psi$ that can be considered as the combination of an electric and magnetic charge, $\boldsymbol\psi=\mathbf e\times \mathbf m$. The number of particles equals the ground state degeneracy on the torus. As all the particles are abelian, the total quantum dimension is $\mathcal D=\sqrt{4}=2$. 

\subsection{Fractional quantum Hall liquids} 
There is a vast literature on the fractional quantum Hall (QH) effect with many pedagogical reviews, see e.g. \cite{Ezawa, Jainbook, Fradkin,TongLectures,qHhierarchies}. 
Here, we want to restrict the discussion to most important properties of quantum Hall liquids, and how they are related to long-range entanglement. 

Quantum Hall liquids form in two-dimensional electron gases -- usually in some variety of high-mobility GaAlAs heterostructures -- at low temperatures and large perpendicular magnetic fields. 
At special fractional filling fractions $\nu=p/q$ --- the filling fraction denotes the ratio of the number of electrons $N_e$ to the number of single-particle states (or flux quanta) $N_\phi$--- the ground states are featureless,  incompressible quantum liquids. 
The bulk excitations are gapped and carry fractional electric charge \cite{Laughlin} --- for an abelian QH state (i.e. a QH state harboring only abelian anyons) at filling $\nu=p/q$, the allowed fractional charges are $n e/q$, $n=0,\ldots, q-1$ \cite{Haldane,Halperin}. 
The edge, on the other hand, is gapless and can be described by a (1+1) dimensional conformal field theory \cite{WenZee}. 
For Laughlin states  with filling fractions $\nu=1/q$, there is a single chiral charged mode flowing along the edge, which is  described by a chiral boson  field. The effective topological field theory contains $q$ abelian quasiparticles labeled by their charge $0,\ldots, \frac{q-1}q$. 
Therefore, the ground state degeneracy on the torus is $q$, and the total quantum dimension is $\mathcal D=\sqrt{q}$ \cite{HaldaneTorus}. 

In the last decade, there was a lot of interest in fractional QH states due to the possibility to realize non-abelian phases that could be utilized for protected quantum computing \cite{TopQuComp}. 
The most promising candidate of a nonabelian QH state is the one at filling fraction $\nu=5/2$ \cite{Willetexp}, which is believed to be accurately described by the Moore-Read state \cite{MooreRead} (or its particle-hole conjugate \cite{antiPfaffian1,antiPfaffian2}). 
Contrary to abelian QH states, it harbors fractional excitations with charge $e/4$ (not $e/2$!), which, in addition, are believed to be so-called Ising anyons \cite{MooreRead}. 
Consequently, the gapless edge theory is described by the Ising conformal field theory\footnote{combined with a chiral boson that describes the charged mode}, which contains two abelian fields (a boson $\mathbf 1$ and a fermion $\mathbf \psi$) and one nonabelian field ($\sigma$) with scaling dimension $\sqrt{2}$. 
It is straightforward to show that the total number of particles in the fermionic state is 6 -- the nonabelian field $\sigma$ can have charges $\pm 1/4$, and the boson and fermion can have charges $0,\, e/2$. 
Thus, the ground state degeneracy on the torus is 6, and the total quantum dimension is $\mathcal D=\sqrt{8}$. 
Note that for nonabelian states, the ground state degeneracy and total quantum dimension indeed provide complementary information. 

\subsection{Entanglement}
Before proceeding to discuss the possibility of measuring the (global) entanglement properties of topological phases of matter, let us first set  notation and introduce some important concepts.  
Assume that the full Hilbert space $\mathcal H$ of our system can be written as a tensor product of two Hilbert spaces $\mathcal H= \mathcal H_A\otimes \mathcal H_B$. We call the corresponding subsystems part $A$ and $B$ in the following. The bipartition of the full space can be done in many different ways, but in actual applications one often chooses a spatial bipartition. For now, let us keep the discussion general. 

Any pure state\footnote{We only consider pure states in this lecture.} in $\mathcal H$ can be expressed in terms of the basis states in $A$ and $B$ as 
\begin{align}\label{Bipartition}
	|\psi\rangle&=\sum_{a,b} c_{a,b} |a_A\rangle |b_B\rangle
\end{align}
where $|a_A\rangle$ ($|b_B\rangle$), $a=1,\ldots, d_A$ ($b=1,\ldots, d_B$) is an orthonormal basis of $\mathcal H_A$ ($\mathcal H_B$). In general, the matrix with entries $c_{a,b}$ is not diagonal. However, we can always find orthonormal bases for $A$ and $B$, which bring it to diagonal form 
\begin{align}\label{SchmidtDecomp}
	|\psi\rangle&=\sum_{i=1}^{n_s} \alpha_i |\phi_i^A\rangle |\phi_i^B\rangle.
\end{align}
This is called the \emph{Schmidt decomposition}\iffindex{Schmidt decomposition} and the Schmidt eigenvalues $\alpha_i$ obey $\alpha_i\geq 0$ and $\sum_{i}\alpha_i^2=1$.\footnote{Note that a complex phase can always be eliminated by incorporating it in the basis states.} The total number of strictly positive eigenvalues $n_s\leq\min(d_A,d_B)$ is called the Schmidt rank. 

 The Schmidt eigenvalues carry information about the entanglement between part $A$ and $B$. If only one eigenvalue is non-zero, i.e.  $\alpha_i=\delta_{i,1}$, the state $|\psi\rangle$ is a product state and called `separable' (un-entangled). 
If several $\alpha_i>0$, the state is called \emph{entangled}. 

The reduced density matrix\iffindex{reduced density matrix} $\rho_A$ of the state $|\psi\rangle$ is obtained by forming the density matrix $|\psi\rangle \langle \psi|$ and tracing out the degrees of freedom in part $B$: 
\begin{align}\label{reduceddensitymatrix}
	\rho_A&=\sum_{b=1}^{d_B} \langle b_B| \left(|\psi\rangle\langle\psi|\right)|b_B\rangle.
\end{align}
This can of course be done  in any basis of $B$, but it becomes particularly simple when using the Schmidt basis $|\phi_i^B\rangle$ from above: 
\begin{align}\label{SchmidtRho}
	\rho_A&=\sum_{i=1}^{n_s} \langle \phi_i^B| \left( 
	\sum_{n,m} \alpha_n\alpha_m|\phi_n^A\rangle \phi_n^B\rangle
	\langle \phi_m^A|\langle \phi_m^B|
	\right)|\phi_i^B\rangle \nonumber\\
	&=\sum_{n=1}^{n_s}\alpha_n^2 |\phi_n^A\rangle\langle\phi_n^A|\nonumber\\
	&=\sum_{n=1}^{n_s}\lambda_n |\phi_n^A\rangle\langle\phi_n^A|
\end{align}
where we used that the Schmidt basis of $B$ is orthonormal. Thus, the eigenvalues of the reduced density matrix are simply the squares of the Schmidt eigenvalues, $\lambda_n=\alpha_n^2$, and the Schmidt basis is  the eigenbasis of $\rho_A$ and $\rho_B$. 

\begin{figure}
    \centering
     \includegraphics[width=0.5\hsize]{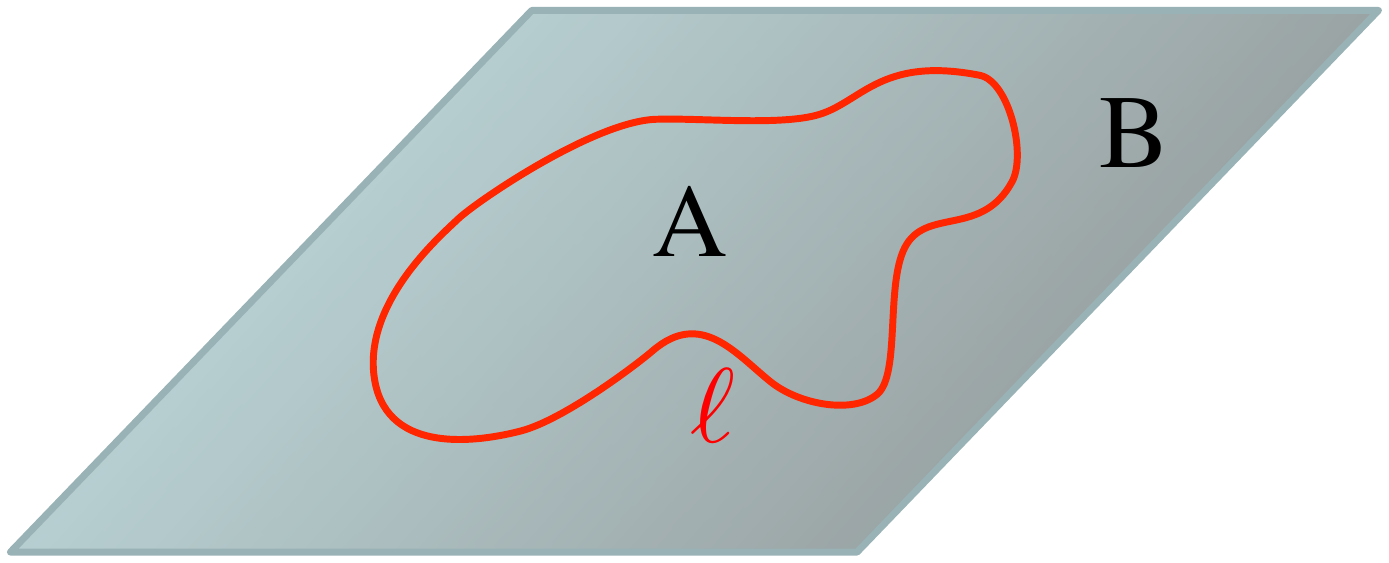}
     \caption{Bipartition of a system into two spatial regions called $A$ and $B$. The entanglement entropy obeys an `area law', i.e. it grows with the circumference $
    \ell$. }\label{fig:entanglement}
\end{figure}

Let us exemplify these concepts with a very simple system of two spin 1/2's. We consider the state
\begin{align}\label{eq:simple ex}
\Psi_\alpha&=(\cos(\alpha)|\uparrow, \uparrow\rangle +\sin(\alpha) |\downarrow, \downarrow\rangle)\nonumber\\
&= (\cos(\alpha)|\uparrow\rangle_A| \uparrow\rangle_B +\sin(\alpha)|\downarrow\rangle_A| \downarrow\rangle_B). 
\end{align}
We have bipartitioned the system so that both $A$ and $B$ contain one of the spins. Eq. \eqref{eq:simple ex} is already in Schmidt form with $\lambda_1= \cos^2(\alpha)$, $ \lambda_2=\sin^2(\alpha)$, and the Schmidt basis is given by $|1_A\rangle=|\uparrow\rangle_A$, $|2_A\rangle=|\downarrow\rangle_A$ (same for part $B$). 
The reduced density matrix of part A is 
\begin{align}
\rho_A =\left( \begin{array}{cc}  \cos^2(\alpha) & 0 \\ 0 & \sin^2(\alpha)  \end{array}\right). 
\end{align}
We see that if $\alpha=0, \frac \pi 2, \pi, \ldots$ then only one of the eigenvalues is nonzero, and the state $\Psi_\alpha$ is separable. On the other hand, if $\alpha=\frac \pi 4, \frac {3\pi} 4, \ldots$ then both eigenvalues are equal and the entanglement between the spins is maximal. In fact,   $\Psi_{\pm \frac \pi 4}$  are two of the four maximally entangled Bell states.

\section{Entanglement entropy}
\iffindex{entanglement entropy}

Let us consider a gapped system that we separate into two spatial regions, called $A$ and $B$ as above. We usually choose a bipartition that is smooth and has no sharp corners -- i.e. the (local) radius of the curvature of the boundary is always much larger than the correlation length $\xi$. We can then define the \emph{entanglement entropy} $S_A$ as the von Neumann entropy of subsystem $A$:
\begin{align}\label{eq:vNeumann}
S_A&=-\mbox{Tr}\left( \rho_A\ln\rho_A\right).
\end{align}
From the diagonal form of the reduced density matrix \eqref{SchmidtRho}, it is easy to see that 
\begin{align}\label{eq:EEschmidt}
S_A&=-\sum_{i=1}^{n_s} \lambda_i \ln \lambda_i=S_B.
\end{align} For generic  states, the entanglement entropy grows as the volume of subsystem $A$. 
 However, for the ground state of a gapped system the leading contribution grows as the  \emph{boundary area} of the system \cite{areaLaw}. This is usually referred to as the \emph{area law}, as the first examples were encountered in (3+1)D quantum gravity \cite{HolographicPrinciple1,HolographicPrinciple2, HolographicPrinciple3}.  

\subsection{Area law and corrections}
\iffindex{area law}

\begin{figure}
    \centering
     \includegraphics[width=.8\hsize]{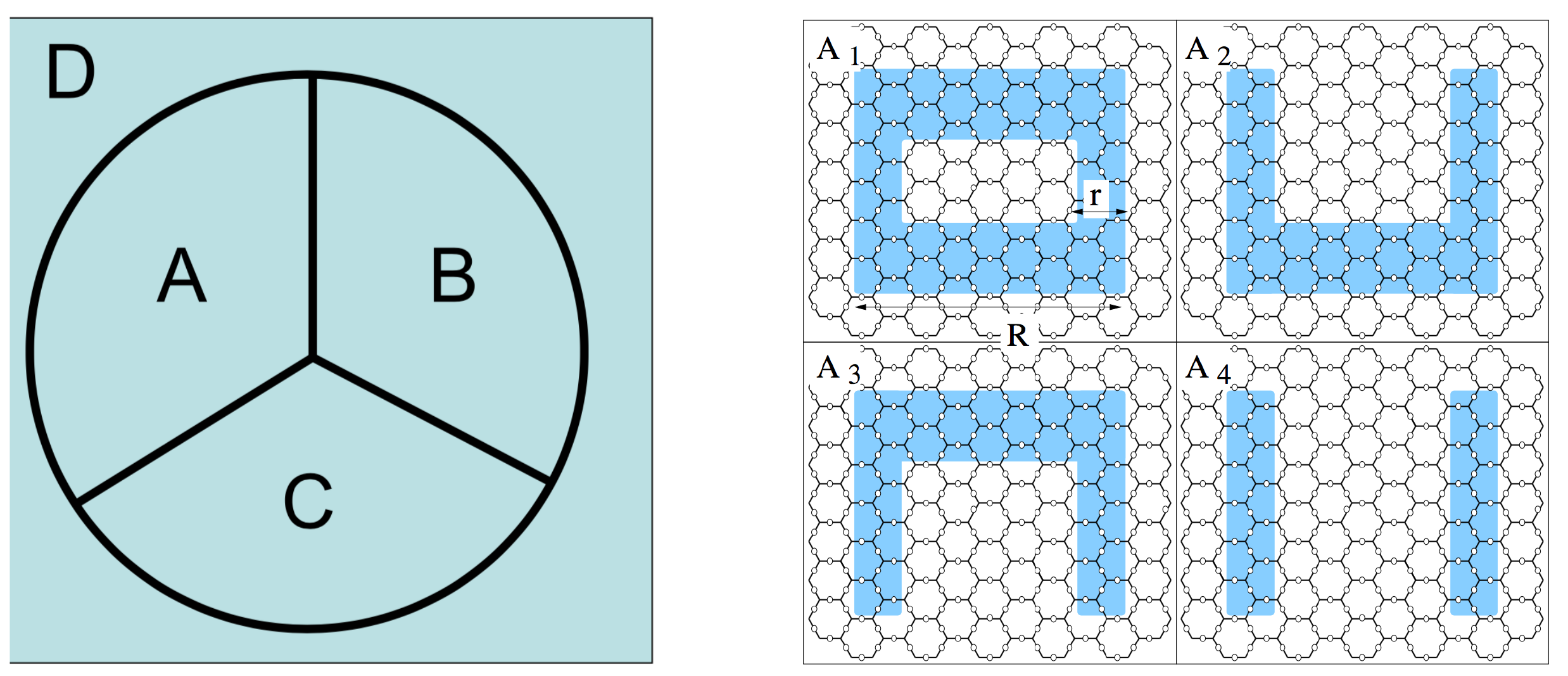}
     \caption{Bipartition schemes by (left) Kitaev and Preskill (picture taken from PRL.96.110404) and (right) Levin and Wen  (picture taken from PRL.96.110405) to extract the topological entanglement entropy without a scaling analysis.  }\label{fig:EEschemes}
\end{figure}

Even though the dominant term in the entanglement entropy grows as the boundary, there are important sub-leading contributions. 
For a simply connected region $A$,  the entanglement entropy has the general form 
\begin{align}
	S_A&=\alpha \ell +b \ln \ell -\gamma	+\mathcal{O}(1/\ell).
\end{align} 
The  logarithmic corrections can arise from corner contributions in  gapless phases or quantum critical points \cite{Cornerterms,EEcriticalpoints}. 
More interesting are the order 1 corrections for gapped phases. Hamma et al. suggested that these corrections to the area law could be a useful tool to determine the topological order of the system \cite{Hamma1}. 
Shortly afterwards, Kitaev and Preskill (KP) \cite{EEtopological1} and Levin and Wen (LW) \cite{EEtopological2} made this more formal and showed that the entanglement entropy has a constant correction that measures the total quantum dimension $\mathcal D$ of the effective TQFT that describes the long-distance behavior of the system. 
The \emph{topological entanglement entropy}\iffindex{topological entanglement entropy} $\gamma=\ln \mathcal D$ can be extracted either by a careful scaling analysis of $S_A$ as a function of the boundary length $\ell$, or by adding/subtracting the entanglement entropy for several, carefully chosen regions (see Fig.~\ref{fig:EEschemes}): 

\begin{align}
KP:\,\,\, \qquad \gamma&\equiv S_A+S_B+S_C-S_{AB}-S_{AC}-S_{BC}+S_{ABC}\nonumber\\
LW:\qquad 2\gamma&\equiv S_1-S_2-S_3 +S_4. 
\end{align} 
The regions and the corresponding subtraction schemes are chosen such that not only the boundary contribution, but also potential corner contributions cancel. 
Note that the LW scheme involves bipartitions that are not simply connected (regions 1 and 4) and for which $\gamma=2\ln \mathcal D$. 

When considering the system on a non-trivial manifold where the ground state is degenerate, one can extract more information from the entanglement entropy than just the total quantum dimension. 
Even though the ground states are degenerate and locally indistinguishable, they do not necessarily have the same entanglement entropy \cite{EEmodularS}. 
In particular, there are special linear combinations for which the topological entanglement entropy $\gamma$ is maximal. As $\gamma$ reduces the entanglement entropy, such states are called \emph{minimally entangled states}. 
Let us, in the following, consider the system on the torus (or alternatively an infinite cylinder, where we can change the topological charge sitting at $\pm \infty$). 
The minimally entangled states can be computed numerically using standard techniques --- e.g.  Monte Carlo techniques as in Ref. \cite{EEmodularS} or tensor network techniques as in Ref. \cite{EEtoporder}.
Using the minimally entangled states, one can then construct the modular $\mathcal S$- and $\mathcal T$-matrices, which contain information about  quasiparticle properties (see e.g. chapter 10 of \cite{CFT} for a detailed discussion on modular invariance, and the definition and properties of the $\mathcal S$- and $\mathcal T$-matrices). 
It was, in fact, conjectured that the modular $\mathcal S$- and $\mathcal T$-matrices contain \emph{all} the information about the topological order \cite{Wenrigid}. 
The numerical method outlined above made it possible to, for the first time, identify the Kalmeyer-Laughlin spin liquid \cite{KalmeyerLaughlin} (a bosonic spin-analog of  Laughlin's QH state \cite{Laughlin}) as the ground state of various (simple) spin-models, see e.g. \cite{KL1,KL2,KL3,KL4}. 

\subsection{Computing the entanglement entropy for the toric code}

 In order to elucidate the discussion above, let us compute the entanglement entropy for the toric code. 
 This was first done by Hamma et al. \cite{Hamma1}; a more general treatment of the entanglement entropy for spin systems can be found in \cite{Hamma2}. 
Without loss of generality, we consider the even/even ground state (i.e. both non-contractible Wilson loop operators \eqref{eq:Wilson} have eigenvalues +1)  on the torus, which can be written as 
\begin{align}\label{eq:TCgroundstate2}
|\Psi\rangle=\frac 1 {\sqrt 2^{N-1}} \prod_{p=1}^{N-1}( 1+B_p)|vac\rangle
\end{align}
with $N$ being the number of plaquettes on the torus and  $|vac\rangle$  the reference state with all spins $\uparrow$. 
The product $\prod_{p=1}^{N-1}(1+B_p)$ reproduces the equal superposition sum of loops without over-counting. Remember that the full product over $p=1,\ldots, N$  counts every loop twice because of the torus constraint
\begin{align}\label{eq:torusconstraint} 
\prod_{p=1}^N B_p=1.
\end{align}
We now bipartition the system along one of the lengths of the torus as shown in Fig.~\ref{fig:TCbipartition}. We can  divide all the plaquettes in three groups -- plaquettes where all the constituting spins are in part $A$ ($B$) are placed in set $I_A$ ($I_B$) and plaquettes that contain  spins in both $A$ and $B$ are placed in the set $I_{AB}$. 

\begin{figure}
    \centering
     \includegraphics[width=.3\hsize]{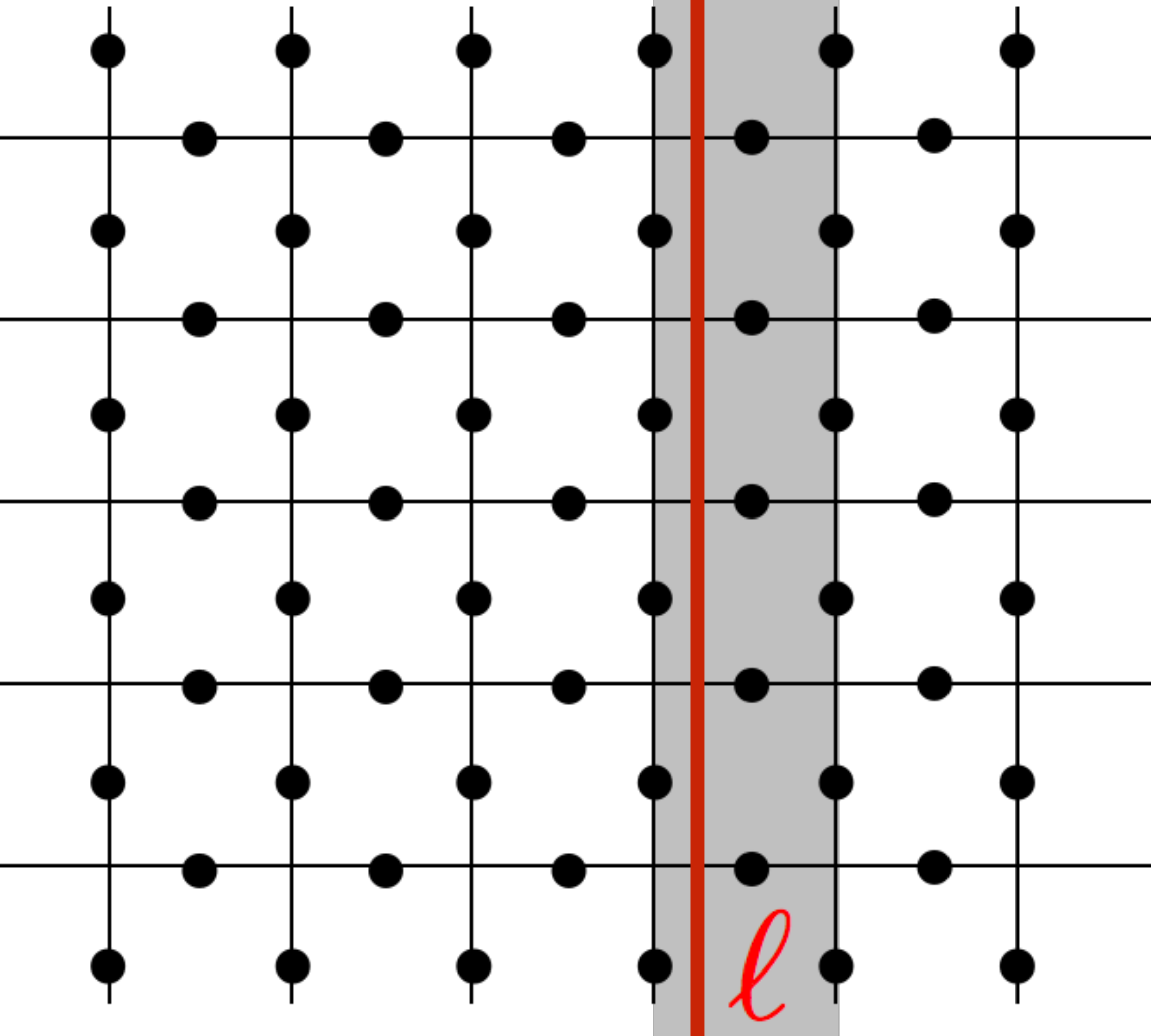}
     \caption{Graphic representation of the spatial bipartition that is used for computing the entanglement entropy. Plaquettes in  $I_{AB}$ are marked in gray. }\label{fig:TCbipartition}
\end{figure}

We can then find the explicit Schmidt decomposition of the ground state \eqref{eq:TCgroundstate2} as 
\begin{align}\label{eq:TCSchmidt1}
|\Psi\rangle&=\frac 1 {\sqrt 2^{N-1}} \prod_{p\in I_{AB}}\!\!'(1+B_p)\prod_{p\in I_{A}}(1+B_p)\prod_{p\in I_{B}}(1+B_p)|vac\rangle_A |vac\rangle_B\nonumber\\
&=\frac {\sqrt{2}^{N_A+N_B}} {\sqrt 2^{N-1}} \prod_{p\in I_{AB}}\!\!'(1+B_p) |\psi\rangle_A |\psi\rangle_B\nonumber\\
&=\frac {1} {\sqrt 2^{\ell-1}} \prod_{p\in I_{AB}}\!\!'(1+B_p) |\psi\rangle_A |\psi\rangle_B. 
\end{align}
Here, $\ell$ denotes the number of plaquettes that have spins both in $A$ and in $B$, i.e. $\ell$ is the length of the boundary, and $N_A$ ($N_B$) is the number of plaquettes in $A$ ($B$). The prime in the product over $I_{AB}$ indicates that the product runs only over $\ell-1$ of the $\ell$ plaquettes -- one of the plaquettes is omitted. 
 This is necessary to avoid over-counting, because the full product $\prod_{p\in I_{AB}}$ generates two loops that lie fully in part $A$ and $B$ respectively, and those were already taken  into account in the products over $I_A(I_B)$. The state $|\psi\rangle_A$ ($|\psi\rangle_B$) is the properly normalized equal weight superposition of all loops that lie entirely in $A$ $(B)$:
\begin{align}\label{eq:basisA}
|\psi\rangle_A &=\frac 1 {\sqrt{2}^{N_A}} \prod_{p\in I_A}(1+B_p)|vac\rangle_A\, ,&|\psi\rangle_B&=\frac 1 {\sqrt{2}^{N_B}} \prod_{p\in I_B}(1+B_p)|vac\rangle_B.
\end{align}

The last line of \eqref{eq:TCSchmidt1} is nothing but the Schmidt decomposition in disguise. 
In order to see this, let us expand the product and define new boundary operators $\mathcal B_{\vec n}^{A(B)}$ as
\begin{align}\label{eq:boundary}
\prod_{p\in I_{AB}}\!\!'(1+B_p)&=\sum_{{n_i=0,1}\atop{i=1,\ldots,\ell-1}} \prod_{i=1}^{\ell-1}B_{p_i}^{n_i} \equiv \sum_{{n_i=0,1}\atop{i=1,\ldots,\ell-1}} \mathcal B_{\vec n}^A\otimes \mathcal B_{\vec n}^B, 
\end{align}
where $\mathcal B_{\vec n}^A$ acts on the boundary spins in part $A$ and  $\mathcal B_{\vec n}^B$ on those in $B$. 
It is straightforward to verify that  $\mathcal B_{\vec n}^A|\psi\rangle_A$ is still a normalized state. 
In addition, $_A\langle \psi| \mathcal B_{\vec m}^A \mathcal B_{\vec n}^A|\psi\rangle_A=0$ if $\vec n\neq \vec m$, because the two states differ in their boundary configuration. 
As the same arguments hold equally well for $B$, we conclude that  the states $\mathcal B_{\vec n}^A|\psi\rangle_A$ and $\mathcal B_{\vec n}^B|\psi\rangle_B$ are nothing but the Schmidt eigenvectors.  From inspection of \eqref{eq:TCSchmidt1} and \eqref{eq:boundary}, it follows  that all the Schmidt values  are identical, $\alpha_i=\frac 1 {\sqrt{2}^{\ell-1}}$.  
Using Eq.~\eqref{eq:EEschmidt} we see that the entanglement entropy is simply
\begin{align}\label{eq:EEtoriccode}
S_A&=-\sum_{i=1}^{2^{\ell-1}}\frac{1}{2^{\ell-1}}\ln \left(\frac 1 {2^{\ell-1}}\right)\nonumber\\
&=(\ell-1)\ln 2,
\end{align}
which gives the correct value of  $\mathcal D=\sqrt{4}$ for the topological entanglement entropy of the toric code. Note that the ground state wave function in any other sector would have given exactly the same result, as changing the sector only amounts to changing the reference state.  

There is also a very simple argument for the reduction of the entanglement entropy by $\ln 2$, based on the graphical representation of the ground state as a loop gas. Loops that reside fully in part $A$ or $B$ cannot contribute to the entanglement between $A$ and $B$, only loops that cross the boundary can. Since the ground state only contains closed loops, there must be an even number of strings crossing the boundary. In the original spin language, this translates to having an even number of  spin-$\downarrow$'s on the boundary. Thus, not $2^\ell$, but only $2^{\ell-1}$ boundary configurations are allowed, which directly leads to the final result of \eqref{eq:EEtoriccode}.

\subsection{R\'enyi entropies and the replica trick}
 \iffindex{R\'enyi entropy}
In numerical simulations, it is often hard to compute the von Neumann entropy, since it requires the full information about the reduced density matrix (or at least sufficiently many of the largest eigenvalues). 
This  information is accessible when doing e.g. exact diagonalization or Density Matrix Renormalization Group calculations, but not when  e.g. using quantum Monte Carlo techniques. 
The latter, however, is amenable to computing R\'enyi entropies
\begin{align}
S_n&=\frac 1{1-n}\ln(\mbox{Tr}[\rho_A^n])\nonumber\\
&=\frac 1{1-n}\ln\left(\sum_i \lambda_i^n\right), 
\end{align}
which in the limit  $n\rightarrow 1$ reproduces the von Neumann entropy \eqref{eq:vNeumann}. 

R\'enyi entropies can be evaluated by introducing $n$ (nearly)\footnote{The only constraint is that they have to match at the boundary to part $A$.} independent replicas of subsystem $B$ -- a tool that was originally introduced in analytical calculations, see e.g. \cite{replica} and \cite{CalabreseCardy1}. 
This technique is, however, also suitable for Monte Carlo simulations \cite{MC1,MC2}. 
The $n$th R\'enyi entropy can be computed by sampling the ratio of two partition functions, $S_n(A)= - \mathcal{Z}[A,n\beta]/\mathcal{Z}^n$, where $\mathcal Z=\mbox{Tr}[\rho_A]$ is the usual partition function of the system, and $\mathcal{Z}[A,n\beta]$ is the partition function of the replicated system, see \cite{MC3} for details. 
By now this trick has been implemented for a variety of Monte Carlo techniques, such as variational Monte Carlo \cite{MCvariational}, continuum-space path-integral Monte Carlo for bosons \cite{MCcontinuumpath} and fermions \cite{MCcontinuumpathF}, determinantal Monte Carlo \cite{MCdeterminantal}, and hybrid Monte Carlo \cite{MChybrid}. 
In principle, one can extract the von Neumann entropies by computing a series of R\'enyi entropies and extrapolating  $n\rightarrow1$. 
However, in practice this is not feasible, and one is usually content with computing the second R\'enyi entropy to extract the topological entanglement entropy. 
The latter was shown to be independent of the R\'enyi index, at least for many of the most relevant topological orders \cite{EEtoric}.

\subsection{Further developments}
The entanglement entropy has become an important tool in many different areas of theoretical condensed matter physics. 
In gapless one-dimensional systems, the entanglement entropy grows logarithmically with the length of part $A$. It was shown that the factor in front of $\ln \ell$ is related to the central charge of the conformal field theory that describes the critical phase \cite{replica,CalabreseCardy}. Computing the entanglement entropy has, by now, become a standard method to identify the nature of critical systems in one dimensions. 

While entanglement is a purely quantum mechanical concept, the entanglement entropy can also be generalized to classical systems by re-interpreting the Schmidt eigenvalues as probabilities of classical configurations in Eq. \eqref{eq:EEschmidt}.
This concept was originally introduced by Castelnovo and Chamon \cite{CastelnovoChamon} for the classical toric code, and later generalized to other classical versions of topologically ordered quantum systems \cite{classicalstringnet,Helmes,classicalphasetransition}. 
The classical entanglement entropy obeys a volume law, but by using the Levin-Wen scheme, one  can again cancel the dominant contributions (volume and boundary) and identify the non-vanishing order 1 contribution. In general, this so-called classical topological entropy is identical to the number of \emph{abelian} fields in the TQFT of the corresponding quantum model \cite{classicalstringnet}.

\section{Entanglement spectrum}
 \iffindex{entanglement spectrum} 

The entanglement entropy is a very powerful tool to detect long-range entanglement in a quantum state, because it reduces the problem to a single quantity, $\gamma >0$, and does not rely on any symmetries. 
However, it does have two shortcomings. 
First, one needs to be able to do a careful scaling analysis or to efficiently implement  the Kitaev-Preskill or Levin-Wen scheme to extract the sub-leading term. 
Second, the total quantum dimension $\mathcal D$ carries far too little information to determine the effective TQFT  for the phase. 
This is exemplified by the two gapped phases of the Kitaev honeycomb model \cite{kitaevhoneycomb} (see  David DiVincenzo lectures) --- a nonabelian phase harboring Ising anyons and an abelian phase that is described by Kitaev's toric code. 
The first  theory has three fields: two abelian fields with quantum dimension one and the Ising anyon with quantum dimension $\sqrt 2$. 
The toric code phase, on the other hand, has four abelian fields.  
The topological entanglement entropy is $\ln 2$ in both cases and, thus, cannot distinguish the two phases. 
However, the full reduced density matrix should (and does in this case! See Ref. \cite{ESKitaev}) contain enough information to distinguish the two phases. 
The only problem that remains is to extract and interpret that information in an effective way.

Li and Haldane proposed the entanglement spectrum as an efficient quantity from which to extract topological information from the reduced density matrix \cite{LiHaldane}. 
They noted that the entanglement entropy can be interpreted as the thermodynamic entropy of a system at temperature $T=1$ with an `entanglement Hamiltonian' $ H_E$ that is defined by $ \rho_A=\exp[- H_E]$. 
Note that this entanglement Hamiltonian is bounded from below by 0 as the eigenvalues of $\rho_A$ satisfy $\lambda_i\leq 1$. 
If the state is weakly entangled, one expects a single  eigenvalue $\lambda_1$ to be close to 1 and the remaining ones to be close to zero. The corresponding \emph{entanglement energies} $\xi_i$ -- defined by $e^{-\xi_i}=\lambda_i$ -- define a gapped system, whose energy gap approaches $\infty$ when the state becomes a product state. 
If the state is strongly entangled, the entanglement spectrum will instead look `gapless'. 

\subsection{Topological properties of the orbital entanglement spectrum}

In their analysis, Li and Haldane used the `orbital cut', which is a numerically efficient way to `mimic' a spatial bipartition in fractional quantum Hall liquids. They found that the entanglement spectrum allows one to identify the conformal field theory that describes the  physics at an edge. This is in fact a quite remarkable finding --  it tells us that we can obtain information about excitations in the system solely from the ground state wave function. This is of course not an uncontroversial result, and it is certainly not true for generic systems. First of all, one may wonder how much a topologically ordered ground state tells us about the excitation spectrum. In general, it tells us nothing. However, if we require it to be the exact ground state of a \emph{gapped} and \emph{local} Hamiltonian, the answer is less straightforward. This is  a very interesting open question that people are still thinking about. To this day, no one has  found a single example of two distinct, gapped, local Hamiltonians that have identical topologically ordered ground state wave functions. While this is far from being a proof, it seems likely that the ground state of a topologically ordered system indeed contains information about the excitations. 

Note that the entanglement spectrum does  not always contain more information about the topological order than the entanglement entropy. A prominent counterexample is the toric code. As all the non-zero Schmidt eigenvalues are identical, the reduced density matrix is a projector, and the corresponding entanglement spectrum is completely flat, with $\mathcal D^2$  eigenvalues at entanglement energy $\xi_i=2\ln\mathcal D$. Thus, it contains only the information about the total quantum dimension, exactly as the entanglement entropy \cite{EEtoric}. The same shortcoming is found for a large set of topologically ordered models, the so-called string-net states \cite{levinwen}. 

For fractional quantum Hall liquids, a number of different bipartitions have proven to be useful to extract information about the system. Quite generically one tries to keep as many symmetries as possible in order to organize the spectrum. 
In the following, we consider quantum Hall liquids on the sphere and choose a gauge so that single-particle states are eigenstates of $L_z$. The good quantum numbers of the ground state are the number of particles $N_e$, the total angular momentum $\mathbf L^2=0$ and its $z$-component $L_z=0$. 
The topological information is encoded in the state counting, i.e. the number of entanglement eigenvalues for given quantum numbers. It is \emph{not} encoded in the entanglement energies -- these change continuously and cannot encode topological (i.e. quantized) information. 

Let us first consider the  `orbital cut' of Li and Haldane. This is actually a bipartition in the angular momentum space -- single-particle orbitals are either placed in part $A$ or part $B$. It mimics a spatial cut, because the angular momentum orbitals are exponentially localized in the latitude. Usually, we choose all single-particle states with $L_z\leq L_z^0$ to be in part $A$ and the remainder in part $B$. The orbital cut breaks the $\mathbf L^2$ symmetry, but keeps particle number and $L_z$ as good quantum numbers. Thus, the reduced density matrix will be block-diagonal with blocks labeled by different values of the tuple $(N_A,L_z^A)$ -- number of particles in $A$ and total angular momentum in $A$. 

If we write a fermionic quantum Hall liquid in Fock space as
\begin{align}
|\Psi\rangle&=\sum_{\lambda} c_\lambda |\lambda\rangle
\end{align}
where $\lambda_1<\lambda_2<\ldots <\lambda_{N}$ denotes the occupied single-particle orbitals, we can directly `read off' the entries of the reduced density matrix from the Fock coefficients. 
Using that in momentum space, each occupation number state factorizes, we can define an orbital entanglement matrix \cite{ESBulkEdge} by
\begin{align}
|\Psi\rangle&=\sum_{i,j} C_{i,j} |\mu_i\rangle |\nu_j\rangle
\end{align}
with $C_{i,j}=c_{\mu_i+\nu_j}$ and $|\mu_i\rangle$ ($|\nu_j\rangle$) containing only  the single-particle orbitals in part $A$ ($B$). 
The reduced density matrix is then given by $\rho_A=CC^\dagger$.

\begin{figure}
    \centering
     \includegraphics[width=\hsize]{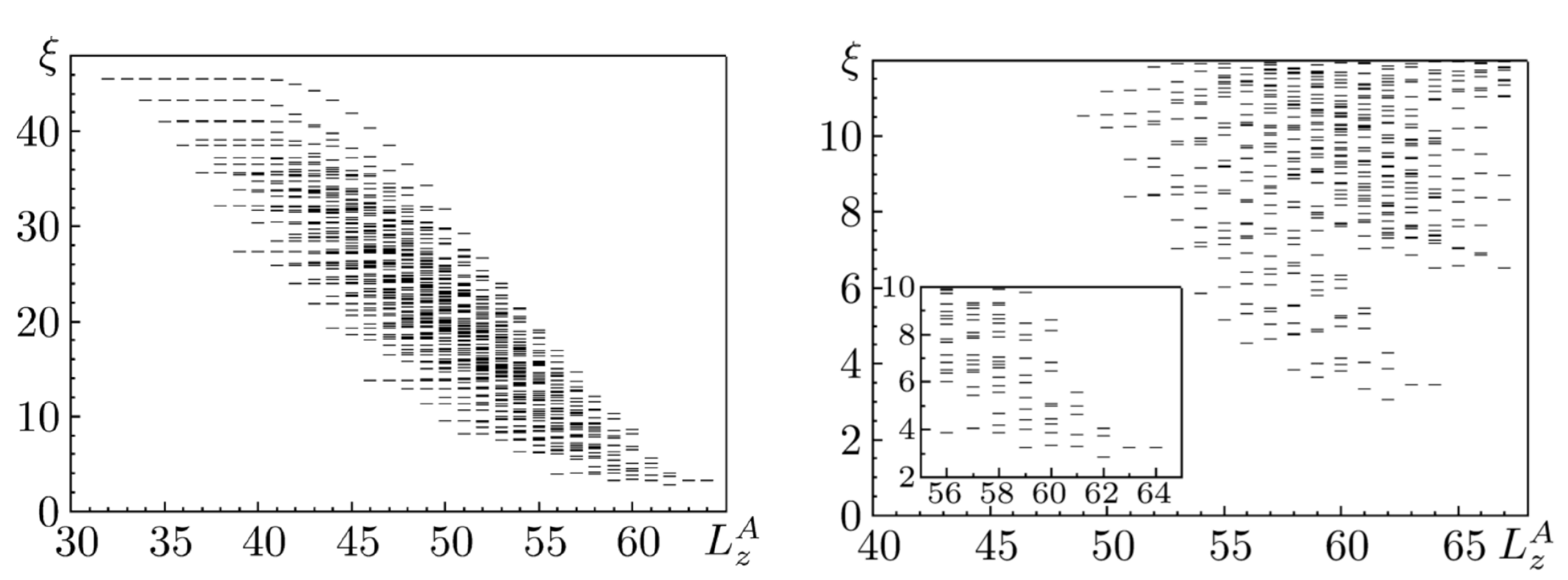}
     \caption{Orbital entanglement spectrum of the Moore-Read model wave function (left) and the exact diagonalization ground state with Coulomb interactions (right). The state counting of the low entanglement energy spectrum of the Coulomb state (see inset of right figure) is identical to that of the Moore-Read state. Higher entanglement energies are separated by a gap (at least for high angular momenta $L_z^A$). (Figures taken from PRL.101.010504)}\label{fig:Oentanglement}
\end{figure}

For quantum Hall \emph{model states}\footnote{Here, we refer to states that are exact zero-energy ground states of certain clustering Hamiltonians, see Ref. \cite{Haldane} for the Laughlin state  and Ref. \cite{GreiterWenWilczek} for the Moore-Read state.}, the rank of the reduced density matrix is usually much(!) smaller than its dimension \cite{LiHaldane}. 
The reason lies in the stringent clustering properties that these states fulfill \cite{Clustering}. 
In a generic state, such as the exact diagonalization ground state of the Coulomb interaction, the dimension of the reduced density matrix usually equals its rank\footnote{The eigenvalues of the reduced density matrix might be very small, but since we already used all the symmetries  to bring the orbital entanglement matrix into block-diagonal form, there is no reason for them to be \emph{exactly} equal to zero.}, as can be seen on the right in Fig. \ref{fig:Oentanglement}. 
However, if the corresponding entanglement spectrum is gapped, the counting of the low-energy spectrum can still contain topological information. 
This was found to be the case for both the Laughlin states and the Moore-Read state. 
The counting of levels at high angular momentum ($L_z^A\leq 64$) and low-entanglement energies ($\xi\lesssim  6$) is identical for the Coulomb and the model state, as can be seen in Fig. \ref{fig:Oentanglement}. 
This can be used to identify the topological order of the Coulomb state, but it also turns out to give direct information on the conformal field theory describing the edge \cite{LiHaldane}. Counting the states for given $N^A $ (as a function of $L_z^A$) exactly reproduces the number of descendent fields (as a function of the `level'\footnote{See e.g. chapter 7.1 of \cite{CFT} for a general discussion on the counting of descendent fields, and chapter 8.1 for the counting in the Ising conformal field theory.})  in the corresponding conformal field theory. 
For Laughlin states, one can show that the orbital entanglement spectrum  in fact contains all the topological information about the phase \cite{ESBulkEdge}. 
One expects that this is true in general, but it has not yet been proven for the more complicated cases such as the Moore-Read state.

\subsection{Other entanglement spectra and their properties}

While the orbital cut is an efficient way to mimic a real-space bipartition, it does not exactly simulate a real-space edge. For strongly correlated spectra, this does not matter too much -- we can still extract the information about the edge spectrum. But for  integer quantum Hall states it does matter. Since they are product states in the momentum space, the orbital entanglement spectrum only has a single eigenvalue at entanglement energy 0. However, we know that the edge theory for the integer quantum Hall liquids is described by a chiral boson. In order to see this chiral boson also in the entanglement spectrum, one needs to make a proper  real-space bipartition. 

Such a  real-space cut is substantially harder to implement numerically than the orbital cut, see  \cite{RSC1,RSC2,RSC3} for a precise definition.\footnote{Note that \cite{RSC2} only reproduces the state counting correctly, but not the entanglement energies.}  
In addition to correctly reproducing the chiral edge theory for an integer quantum Hall state, it also has the  feature that not only its state counting, but also its entanglement energies, are correctly predicted by the conformal field theory (in the thermodynamic limit) \cite{RSC1}. This correspondence between entanglement spectra and edge spectra has also been shown in  more general contexts by Qi, Katsura, and Ludwig \cite{bulkEdge1} and Swingle and Senthil \cite{Swingle}. 
The real-space entanglement spectrum is believed to contain exactly the same information about the topological order of the ground state as the orbital entanglement spectrum. This was shown explicitely for the Laughlin states, but is expected to hold in general \cite{ESBulkEdge,RSC1}.  

Another important bipartition is the `particle cut', where a number of particles are integrated out \cite{PESHaque1,PESHaque2,PES}. 
This bipartition is again relatively easy to implement numerically. 
Just as for the orbital cut, we can define a particle entanglement matrix $\mathbf P$ by writing the ground state in Fock space as 
\begin{align}
|\Psi\rangle&=\sum_{i,j} P_{i,j} |\mu_i\rangle |\nu_j\rangle,
\end{align}
where $\mu_j$ ($\nu_j$) are Fock states of $N_A$ ($N-N_A$) particles in the full single-particle orbital space, and $P_{i,j}=c_{\mu_i+\nu_j}$. The reduced density matrix is again given by $\mathbf P \mathbf P^\dagger$. 
The particle cut has one additional quantum number compared to the orbital and real space cut, namely the total angular momentum $\mathbf L^2$. Thus, all entanglement eigenvalues can be organized in multiplets, and the spectrum is flat. The state counting --- i.e. the number of levels as a function of $N_A$ and $L_z^A$ --- is identical to that of the real space cut (see for instance \cite{RSC2} for a simple proof) and greater or equal  than that of the orbital cut. When restricting to high $L_z^A$, one can show that all the three spectra have the same state counting \cite{ESBulkEdge}, but in the intermediate $L_z^A$ regime the orbital entanglement has fewer levels.

The particle entanglement spectrum provides information about the \emph{bulk excitations} of the system \cite{PES}. 
This is simplest to see from a real-space perspective. Tracing out a number of particles does not change the vanishing properties of the remaining particles. Now we need to remember that several model states, such as Laughlin or Moore-Read, can be defined by their vanishing properties. The densest state obeying the vanishing conditions defines the ground state, and less dense states are interpreted as quasihole states \cite{ReadRezayi,POZ1,POZ2}. This implies that the eigenstates of the reduced density matrix are the quasihole states of the QH liquid with less (namely $N_A$) particles, but the full flux of the original ground state $N_\phi$. This analogy also gives us an upper bound on the number of low-lying entanglement energies in the particle entanglement spectrum -- it is bounded by the number of quasihole states of $N_A$ particles in flux $N_\phi$.  The latter number is well-known from the conformal field theory description \cite{MooreRead}. Numerical simulations indicate that this bound is always reached, but a theoretical proof of this observation is so far lacking, see  Ref.~\cite{Ardonne} for more details.

As the particle entanglement spectrum is straight-forward to implement numerically (in contrast to the real-space entanglement spectrum), and has a nice physical interpretation (in contrast to the orbital entanglement spectrum), it can be generalized to other systems. In particular, it has been instrumental in determining the topological order in fractional Chern insulators \cite{fractionalCI}. A more detailed discussion on the various entanglement spectra and their applications can be found in the lecture notes by Regnault \cite{EntSpectroscopy}.

\subsection{Further developments}

Entanglement spectra are not only relevant for strongly correlated states, but can also provide valuable information for noninteracting topological states. 
For the latter, it is  sufficient to look at the single-particle entanglement spectrum, as the entanglement spectra for more particles can be constructed from the single-particle one \cite{Peschel}. 
In fact, the topological information of a real-space partition can be organized in a very clear way by plotting the spectrum of the correlation matrix $\langle gs| c_i^\dagger c_j |gs\rangle$, where $i$ and $j$ are restricted to sites in part $A$. 
It has the same information as the single-particle entanglement spectrum: the entanglement energies $\epsilon_l$ can be expressed in terms of the eigenvalues of the correlation matrix $\xi_l$ as $\epsilon_l=\ln(1-\xi_l)-\ln \xi_l$. 
In addition, it is directly related to the spectrally flattened Hamiltonian of the physical system. 
As such, the entanglement spectrum provides a very simple way to identify non-interacting topological phases \cite{fidkowski}, and was even suggested to be a more robust measure of topology than the edge mode spectrum itself \cite{Vishwanath}.

The entanglement spectrum was also proposed as a useful tool in one-dimensional systems, both as a  way to comprehensively classify gapped topological phases \cite{ES1Dtopological} and to extract information about the conformal field theory describing critical points \cite{Laeuchli}.

\end{document}